\documentclass{iaus}
\usepackage{graphicx}

\title[Progress report on solar age calibration]
{Progress report on solar age calibration}

\author[Houdek \& Gough]   
{G. Houdek$^1$ \and D.O. Gough$^{1,2}$}%

\affiliation{$^1$Institute of Astronomy, Madingley Road,
Cambridge CB3\,0HA, UK \break email: hg@ast.cam.ac.uk\\[\affilskip]
$^2$Department of Applied Mathematics and Theoretical Physics,
Wilberforce Road, \break Cambridge CB3\,0WA, UK 
\break email: douglas@ast.cam.ac.uk}

\pubyear{2008}
\volume{252}  
\pagerange{1--8}
\date{?? and in revised form ??}
\jname{The Art of Modelling Stars in the 21st Century}
\editors{Licai Deng, K.L. Chan \& C. Chiosi, eds.}
\begin{document}

\maketitle

\begin{abstract}
We report on an ongoing investigation into a seismic calibration of
solar models designed for estimating the main-sequence age and a measure 
of the chemical abundances of the Sun. Only modes of low degree are employed,
so that with appropriate modification the procedure could be applied
to other stars. We have found that, as has been anticipated, a separation
of the contributions to the seismic frequencies arising from the
relatively smooth, glitch-free, background structure of the star and
from glitches produced by helium ionization and the abrupt gradient
change at the base of the convection zone renders the procedure more
robust than earlier calibrations that fitted only raw frequencies to
glitch-free asymptotics. As in the past, we use asymptotic analysis to
design seismic signatures that are, to the best of our ability,
contaminated as little as possible by those uncertain properties of
the star that are not directly associated with age and chemical composition.
The calibration itself, however, employs only numerically computed
eigenfrequencies. It is based on a linear perturbation from a reference
model. Two reference models have been used, one somewhat younger, the other 
somewhat
older than the Sun. The two calibrations, which use BiSON data, are
more-or-less consistent, and yield a main-sequence age
$t_\odot=4.68\pm0.02\,$Gy, coupled with a formal initial 
heavy-element abundance $Z=0.0169\pm0.0005$. The error analysis has not 
yet been completed, so the estimated precision must be taken with a 
pinch of salt.
\end{abstract}

\firstsection 

\section{Introduction}

The only way by which the age of the Sun can be estimated to a useful degree
of precision is by accepting the basic tenets of solar-evolution theory and
measuring those aspects of the structure of the Sun that are predicted by
the theory to be indicators of age. The structure measurements must be
carried out seismologically, and evidently one expects greatest reliability
of the results when all the available helioseismic data are employed.
However, the most relevant modes are those of lowest degree, because it is
they that penetrate most deeply into the energy-generating core where the
relic helium-abundance variation records the integrated history of nuclear
transmutation. Moreover, it is also only they that can be measured in 
other stars. Therefore, there has been some interest in calibrating
theoretical stellar models using only low-degree modes. The prospect was 
first discussed in detail by 
\cite[Christensen-Dalsgaard (1984, 88)]{jcd84, jcd88}, \cite{ulrich86} 
and \cite{dog87}, although prior to that it had already been pointed out
that the helioseismic frequency data that were available at the time 
indicated that either the initial helium abundance $Y_0$ or the age $t_\odot$,
or both, are somewhat greater than the generally accepted values
\cite[(Gough 1983)]{dog83}, an inference which is consistent with our 
present findings. Subsequent, more careful, calibrations were carried out
by \cite{guenther89}, \cite{dog-nov90}, \cite{guenther-demarque97}, 
\cite{weiss-schlattl98}, \cite{wd99}, \cite{dog01} and
Bonanno, Schlattl \& Patern\`o (2002). Not all of them 
addressed the influence of uncertainties
in $Y_0$ on the determination of $t_\odot$.

As a main-sequence star ages, helium is produced in the core, increasing
the mean molecular mass $\mu$, preferentially at the centre, and thereby
reducing the sound speed. The resulting functional form of the sound speed
$c(r)$ depends not only on age $t_\odot$ but also on the relative
augmentation of $\mu(r)$, which itself depends on the initial absolute
value of $\mu$, and hence on $Y_0$. \cite{dog01} tried to separate these 
two effects using the degree dependence of the small separation
$d_{n,l}=3(2l+3)^{-1}(\nu_{n,l}-\nu_{n-1,l+2})$ of cyclic frequencies
$\nu_{n,l}$, where $n$ is order and $l$ is degree. This is possible, in
principle, because modes of different degree and similar frequency sample the
core differently. However, that difference is subtle, and the sensitivity 
to the relatively fine distinction between the effects of $t_\odot$ and
$Y_0$ on the functional form of $c(r)$ in the core is low.
Consequently the error in the calibration produced by errors in the observed
frequency data is uncomfortably high.

This lack of sensitivity can be overcome by using, in addition to 
core-sensitive seismic signatures, the relatively small oscillatory component 
of the eigenfrequencies induced by the sound-speed glitch associated with
helium ionization \cite[(Gough 2002)]{dog02}, whose amplitude is close to
being proportional to helium abundance $Y$ 
\cite[(Houdek \& Gough 2007a)]{hgdog07a}. The neglect of that component
in the previously employed asymptotic signature $d_{n,l}$ had not only omitted
an important diagnostic of $Y$ but had also imprinted an oscillatory
contamination in the calibration as the limits $(k_1, k_2)$, where
$k=n+\frac{1}{2}l$, of the adopted mode range was varied 
\cite[(Gough 2001)]{dog01}. It therefore behoves us to decontaminate the core
signature from glitch contributions produced in the outer layers of the star
(from both helium ionization and the abrupt variation at the base of the 
convection zone, and also from hydrogen ionization and the superadiabatic
convective boundary layer immediately beneath the photosphere). To this end
a helioseismic glitch signature has been developed by
\cite{hgdog07a}, from which frequency contributions
$\delta\nu_{n,l}$ can be computed and subtracted from the raw frequencies
$\nu_{n,l}$ to produce effective glitch-free frequencies $\nu_{{\rm s}n,l}$
to which a glitch-free asymptotic formula (\ref{eq:asymp}) can be fitted.
The solar calibration is then accomplished by interpolating the theoretical 
seismic
signatures computed on a grid of solar models to the observations, using
a standard grid to compute derivatives with respect to $t_\odot$ and $Y_0$,
and a carefully computed reference solar model designed to be close to the
Sun. The result of the first preliminary calibration by this method, using
BiSON data, has been reported by \cite{hgdog07b}. Here we enlarge on our
discussion of the analysis, and we augment our results with a calibration 
based on a second reference solar model.
\vspace{-5mm}
\begin{figure}
\centering
\includegraphics[height=.30\textheight]{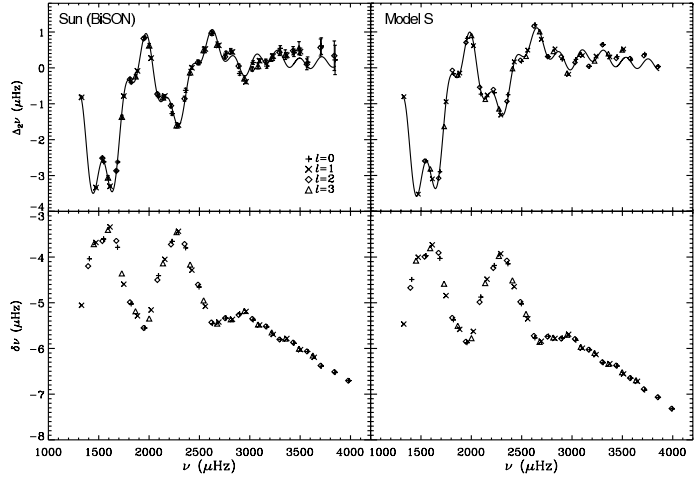}
\caption{Top left: The symbols (with error bars obtained 
under the assumption that the raw frequency errors are independent) 
represent second differences, $\Delta_2\nu$, of low-degree solar
frequencies from BiSON.
Top right: The symbols are second differences $\Delta_2\nu$
of adiabatic pulsation eigenfrequencies of solar 
Model S of \cite{jcd96}.
The solid curves in both panels are the diagnostics\,
(\ref{eq:delnu}) -- (\ref{eq:secdiff}), 
whose eleven parameters have been adjusted to fit the data optimally.
Bottom: The symbols denote contributions $\delta\nu$ to the frequencies  
produced by the acoustic glitches 
of the Sun (left panel) and Model S (right panel).}
\end{figure}
\section{The seismic diagnostic and calibration method}
Any abrupt variation in the stratification of a star (relative to the scale of 
the inverse radial wavenumber of a seismic mode of oscillation), which here
we call an acoustic glitch, induces an oscillatory component in the spacing of 
the cyclic eigenfrequencies of seismic modes.
Our interest is principally in the glitch caused by the 
depression in the first adiabatic exponent 
$\gamma_1=(\partial {\ln p}/\partial{\ln\rho})_s$ (where $p$, $\rho$ and $s$ 
are pressure, density and specific entropy) caused by helium ionization, which
imparts a glitch in the sound speed $c(r)$. The deviation 
\vspace{-2mm}
\begin{equation}
\delta\nu_i:=\nu_i - \nu_{{\rm s}i}\,,
\label{eq:nudiff}
\end{equation}
where $i:=(n, l)$, of the eigenfrequency $\nu_i$ from the corresponding 
frequency $\nu_{{\rm s}i}$ 
of a similar smoothly stratified star is the indicator of $Y$ that 
we use in conjunction with the indicators of core structure to determine the 
main-sequence age. 

Approximate expressions for the frequency contributions $\delta\nu_i$ arising 
from acoustic glitches in solar-type stars were recently 
presented by \cite{hgdog07a}.  Here we improve them by adopting the 
appropriate Airy functions Ai$(-x)$ that are used as comparison 
functions in the JWKB approximations to the oscillation eigenfunctions,  
as did \cite{hgdog07b}.
The complete expression for $\delta\nu_i$ is then given by 
\begin{equation}
\delta\nu_i=\delta_\gamma\nu_i+\delta_{\rm c}\nu_i\, ,
\label{eq:delnu}
\end{equation}
\vspace{-4mm}
where
\begin{eqnarray}
\delta_\gamma\nu&=&-\sqrt{2\pi}A_{\rm II}\Delta^{-1}_{\rm II}
\left[\nu+\textstyle\frac{1}{2}(m+1)\nu_0\right]\cr
&&\hspace{-8pt}
\times\Bigl[\tilde\mu\tilde\beta\int_0^T\kappa^{-1}_{\rm I}
{\rm e}^{-(\tau-\tilde\eta\tau_{\rm II})^2/2\tilde\mu^2\Delta^2_{\rm II}}|x|^{1/2}
|{\rm Ai}(-x)|^2\,{\rm d}\tau\cr
&&\;\;+\;\int_0^T\kappa^{-1}_{\rm II}
{\rm e}^{-(\tau-\tau_{\rm II})^2/2\Delta^2_{\rm II}}|x|^{1/2}
|{\rm Ai}(-x)|^2\,{\rm d}\tau\Bigr]
\label{eq:delgamnu}
\end{eqnarray}
arises from the variation in $\gamma_1$ induced by helium ionization, and
\begin{eqnarray}
\delta_{\rm c}\nu&\simeq&A_{\rm c}\nu_0^3\nu^{-2}
   \left(1+1/16\pi^2\tau_0^2\nu^2\right)^{-1/2}\cr
&\times&\hspace{-3pt}\left\{\cos[2\psi_{\rm c}+\tan^{-1}(4\pi\tau_0\nu)]
      \!-\!(16\pi^2\tilde{\tau}_{\rm c}^2\nu^2\!+\!1)^{1/2}
\right\}\, 
\label{eq:delcnu}
\end{eqnarray}
arises from the acoustic glitch at the base of the convection zone resulting
from a near discontinuity (a true discontinuity in theoretical models using
local mixing-length theory with a non-zero mixing length at the lower 
boundary of 
the convection zone) in the second derivative of density.
Here, $m=3.5$ is a constant, being a representative polytropic index in
the expression for the approximate effective phase $\psi$ appearing in the 
argument of the Airy function, and $\tilde\beta$, $\tilde\eta$ and 
$\tilde\mu$ are
constants of order unity which account for the relation between the acoustic 
glitches caused by the first and second stages of ionization of helium 
(\cite[Houdek \& Gough 2007a]{hgdog07a}); $\tau$ is acoustic depth beneath 
the seismic surface of the star, and $T\simeq1/2\nu_0$ is the total acoustic 
radius of the star; 
$\Delta_{\rm II}$ and $\tau_{\rm II}$ are respectively the acoustic width of 
the glitch and its acoustic depth beneath the seismic surface. The 
argument of the Airy function is $x={\rm sgn}(\psi)|3\psi/2|^{2/3}$, where 
\begin{equation}
\psi(\tau)=\kappa\omega\tilde\tau-(m+1)\cos^{-1}[(m+1)/\omega\tilde\tau]
\hspace{23pt}{\rm if}\ \tilde\tau>\tau_{\rm t}\,,
\end{equation}
\vspace{-2mm}
and
\begin{equation}
\psi(\tau)=|\kappa|\omega\tilde\tau-(m+1)\ln[(m+1)/\omega\tilde\tau+|\kappa|]
\hspace{10pt}{\rm if}\ \tilde\tau\le\tau_{\rm t}\,,
\end{equation}
in which 
$\tilde\tau=\tau+\omega^{-1}\epsilon_{\rm II}$, with $\omega=2\pi\nu$, and 
$\tau_{\rm t}$ is the location of the upper turning point of the mode; also
$\kappa(\tau)=[1-(m+1)^2/\omega^2\tilde\tau^2]^{1/2}\,,$
and $\kappa_{\rm I}=\kappa(\tilde\eta\tau_{\rm II})$ and 
$\kappa_{\rm II}=\kappa(\tau_{\rm II})$. In addition
\vspace{-2mm}
\begin{equation} 
\psi_{\rm c}=\kappa_{\rm c}\omega\tilde\tau_{\rm c}
-(m+1)\cos^{-1}\left[(m+1)/\tilde\tau_{\rm c}\omega\right]+\pi/4,\,
\end{equation}
where 
$\kappa_{\rm c}=\kappa(\tau_{\rm c})$ and 
$\tilde\tau_{\rm c}=\tau_{\rm c}+\omega^{-1}\epsilon_{\rm c}$.

The seven coefficients 
$\eta_\alpha=(A_{\rm II}$, $\Delta_{\rm II}$, $\tau_{\rm II}$, 
$\epsilon_{\rm II}$, $A_{\rm c}$, $\tau_{\rm c}$, $\epsilon_{\rm c}),\, 
\alpha=1,...,7,$ are found by fitting the second difference 
\vspace{-3mm}
\begin{equation}
\Delta_{2i}\nu\equiv\nu_{n-1,l}-2\nu_{n,l}+\nu_{n+1,l}
\simeq\Delta_{2i}(\delta_\gamma\nu+\delta_{\rm c}\nu)+\sum_{k=0}^3a_k\nu_i^{-k}\equiv g_i(\nu_j;\eta_\alpha)
\label{eq:secdiff}
\vspace{-1mm}
\end{equation}
to the corresponding observations by minimizing
\begin{equation}
E_{\rm g}=(\Delta_{2i}\nu-g_i)C^{-1}_{\Delta ij}(\Delta_{2j}\nu-g_j)
\label{eq:minsecdiff}
\end{equation}
using the value of $\nu_0$ obtained by fitting to (\ref{eq:asymp}),
where $C^{-1}_{\Delta ij}$ is the $(i,j)$ element of the inverse of the 
covariance matrix {C$_\Delta$} of the observational errors in 
$\Delta_{2i}\nu$, computed, perforce, under the assumption that the 
errors in the frequency data $\nu_i$ are independent. The last term in 
equation\,(\ref{eq:secdiff}) approximates smooth contributions
arising, in part, from wave 
refraction in the stellar core, from hydrogen ionization and from the 
superadiabaticity of the upper boundary layer of the convection zone, 
introducing four more fitting coefficients 
$a_k=\eta_\alpha$, $k=0,...,3$,\, $\alpha=8,...,11$.
The covariance matrix $C_{\eta\alpha\gamma}$ of the errors in $\eta_\alpha$
were established by Monte Carlo simulation.

The outcome of the fitting to the BiSON data (\cite[Basu et al. 2007]{basu07}) 
and to the adiabatically computed eigenfrequencies of solar  Model S 
(\cite[Christensen-Dalsgaard et al. 1996]{jcd96}) is displayed in Figure\,1: 
the upper panels display the second differences, 
together with the fitted formula (\ref{eq:secdiff}), the lower panels 
display the corresponding contributions $\delta\nu_i$ to the frequencies of 
oscillation from the acoustic glitches.
All the frequencies displayed in the figure have been used in equation
(\ref{eq:minsecdiff}) for fitting (\ref{eq:secdiff}).

To the resulting glitch-free frequencies $\nu_{{\rm s}i}$, derived from 
equation (\ref{eq:nudiff}), of both the solar observations 
and the eigenfrequencies of the reference solar model, is fitted the 
asymptotic expression
\vspace{-2mm}
\begin{equation}
\nu_{{\rm s}i}\!\sim\!(n+{\textstyle\frac{1}{2}}\,l+\hat\epsilon)\nu_0
-\frac{AL^2\!\!-\!B}{\nu_{{\rm s}i}}\,\nu^2_0
-\frac{CL^4\!\!-\!DL^2\!+\!E}{\nu_{{\rm s}i}^3}\,\nu^4_0
-\frac{FL^6\!\!-\!GL^4\!+\!HL^2\!-\!I}{\nu_{{\rm s}i}^5}\,\nu^6_0
\equiv s(\nu_{{\rm s}i};\xi_\beta)\,,
\label{eq:asymp}
\end{equation}
by minimizing $(\nu_{{\rm s}i}-s_i)C^{-1}_{{\rm s}ij}(\nu_{{\rm }j}-s_j)$,
where $L^2=l(l+1)$ and C$_{\rm s}$ is the covariance matrix of the 
observational errors in $\nu_{{\rm s}i}$, from which we obtain both the 
coefficients $\xi_\beta=(\nu_0, ~\hat\epsilon,~A, B, C, D, E, F, G, H, I)$,
$\beta=1,...,11$, and the covariance matrix $C_{\xi\beta\delta}$ of the errors.
Following \cite{dog01}, we carry out this fitting in 
the frequency range given by $k_1\le k\le k_2$, where 
$k=n+\frac{1}{2}l$ and $0\le l\le3$, and we vary
$k_1$ and $k_2$. Each of the parameters $\xi_\beta$
represents an integral of a function of the equilibrium stratification. 
The integrals $A, C$ and $F$ are of particular importance to our analysis, 
because $C$ and $F$ are dominated by conditions in the core, and, although 
the contributions to $A$ from the core and the rest of the star are 
roughly equal in magnitude (and potentially have opposite signs), 
the latter is relatively insensitive to $t_\odot$ and $Y_0$. The integrands 
in the remaining integrals are either more evenly distributed throughout 
the Sun or are concentrated near the surface.


We have carried out age calibrations using combinations of the parameters 
\begin{equation}
\zeta_\alpha=(A,C,-\delta\gamma_1/\gamma_1),\qquad \alpha=1,2,3\,,  
\label{eq:xi}
\end{equation}
where $-\delta\gamma_1/\gamma_1=A_{\rm II}/\sqrt{2\pi}\nu_0\Delta_{\rm II}$ is
a measure of the maximum depression in $\gamma_1$ in the second helium 
ionization zone. 
Presuming, as is normal, that the reference model is parametrically close 
to the Sun, we consider the reference value 
$\zeta^{\rm r}_\alpha$ to be approximated by a two-term Taylor 
expansion of $\zeta_\alpha$ about the value $\zeta^\odot_\alpha$ of
the Sun:
\vspace{-1mm}
\begin{equation}
\zeta^{\rm r}_\alpha=\zeta^{\odot}_\alpha
   -\left(\frac{\partial\zeta_\alpha}{\partial t_\odot}\right)_{\!\!Z}\Delta\,t_\odot
   -\left(\frac{\partial\zeta_\alpha}{\partial Z}\right)_{\!\!t_\odot}\Delta Z
   +\epsilon_{\zeta\alpha}\, ,
\end{equation}
where $\Delta\,t_\odot$ and $\Delta Z$ are the deviations of age $t_\odot$ and
initial heavy-element abundance $Z$ from the reference model, and 
$\epsilon_{\zeta\alpha}$
are the formal errors in the calibration parameters whose covariance
matrix $C_{\zeta\alpha\beta}$ can be derived from $C_{\xi\beta\delta}$
and $C_{\eta\alpha\gamma}$. A (parametrically local) maximum-likelihood 
fit then leads to the following set of linear equations: 
\vspace{-1mm}
\begin{equation}
H_{\alpha j}C^{-1}_{\zeta\alpha\beta}H_{\beta k}\Theta_{0k}=
H_{\alpha j}C^{-1}_{\zeta\alpha\beta}\Delta_{0\beta}\,,
\label{eq:calib1}
\vspace{-1pt}
\end{equation}
in which $\Theta_k=(\Delta t_\odot, \Delta Z)+\epsilon_{\Theta k}=
\Theta_{0k}+\epsilon_{\Theta k}$, $k=1,2$, is the solution vector subject 
to (correlated) errors $\epsilon_{\Theta k}$, 
$\Delta_\beta=\zeta^\odot_\beta-\zeta^{\rm r}_\beta+\epsilon_{\zeta\beta}
=\Delta_{0\beta}+\epsilon_{\zeta\beta}$, and the partial derivatives 
$H_{\alpha j}=[(\partial\zeta_\alpha/\partial t_\odot)_Z,
(\partial\zeta_\alpha/\partial Z)_{t_\odot}]$, $j=1,2$.

A similar set of equations is obtained for the formal errors 
$\epsilon_{\Theta k}$: 
\begin{equation}
H_{\alpha j}C^{-1}_{\zeta\alpha\beta}H_{\beta k}\epsilon_{\Theta k}=
H_{\alpha j}C^{-1}_{\zeta\alpha\beta}\epsilon_{\zeta\beta}\,,
\label{eq:calib2}
\end{equation}
from which the error covariance matrix 
$C_{\Theta kq}=\overline{\epsilon_{\Theta k}\epsilon_{\Theta q}}$ can be 
computed from $C_{\zeta\alpha\beta}$.

The partial derivatives $H_{\alpha j}$ were obtained from the
two sets of five calibrated evolutionary models for the Sun that
were used in a similar calibration by \cite{hgdog07b},
computed with the evolutionary programme by \cite{jcd82}, and  
adopting the Livermoore equation of state and the OPAL92 opacities. 
One set of models has a constant value for the heavy-element abundance 
$Z=0.02$ but varying age; the other has constant age but varying $Z$. 
Note that, for prescribed relative abundances of heavy elements,
the condition that the luminosity and radius of the Sun agree with
observation defines a functional relation between $Y_0, Z$ and $t_\odot$.
The values of the partial derivatives $H_{\alpha j}$ are listed in Table\,1.
\vspace{-4mm}
\begin{table}
\centering
\caption{Partial derivatives $H_{\alpha j}$ obtained from two sets of
calibrated evolutionary models for the Sun. Values with respect
to age $t_\odot$ are in units of Gy$^{-1}$.}
\begin{tabular}{cccccc}
\noalign{\smallskip}
\noalign{\smallskip}
\hline
 \hfil$(\partial A/\partial t_\odot)_Z$\hfil&\hfil$(\partial A/\partial Z)_{t_\odot}$\hfil&
 \hfil$(\partial C/\partial t_\odot)_Z$\hfil&\hfil$(\partial C/\partial Z)_{t_\odot}$\hfil&
 \hfil$[\partial(-\delta\gamma_1/\gamma_1)/\partial t_\odot]_Z$\hfil&
 \hfil$[\partial(-\delta\gamma_1/\gamma_1)/\partial Z]_{t_\odot}$\hfil\\
\hline
\,-0.0469\,&\,-0.584\,&\,0.677\,&\,36.8\,&\,-0.00656\,&\,0.442\,\\
\hline
\end{tabular}
\end{table}
\section{Results}
To illustrate the effect of taking $\delta\nu_i$ into account, we compare 
in Figure\,2 a first assessment 
of $A(k_1,k_2)$ using the glitch-free frequencies $\nu_{{\rm s}i}$ (left panel) 
with that obtained from the raw frequencies $\nu_{i}$ (right panel). Recall
that $A$ represents a functional of the equilibrium structure of the star, and
should not vary with $k_1$ and $k_2$. The range of values for $A$ is the
lower for $\nu_{{\rm s}i}$, as we had anticipated. We believe that the upturn 
of $A$ for low values of $k_1$ in the left panel of the figure is a result 
of the failure of the asymptotic formula (\ref{eq:delnu})--(\ref{eq:delcnu}) 
when $\nu_i$ is low. The dipping of $A$ at high values of $k_1$ and low 
values of $k_2$ occurs because the frequency range is too small for 
a reliable 
determination of the fitting coefficients $\xi_\beta$. We therefore adopt
intermediate values for $k_1$ and high values for $k_2$, for which $A$ is 
insensitive to the selected frequency range.
\begin{figure}
  \includegraphics[height=.22\textheight]{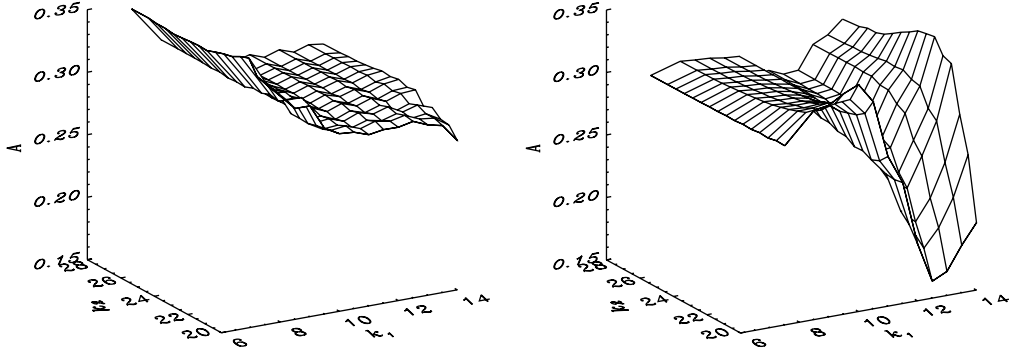}
  \caption{Asymptotic fitting coefficient $A$ (see equation \ref{eq:asymp}) as
           functions of $k_1$ and $k_2$ ($k=n+\frac{1}{2}l$). Results
           are shown for fitting (\ref{eq:asymp}) to the glitch-free frequencies
           $\nu_{{\rm s}i}$ (left panel) and to the raw frequencies $\nu_i$ 
           (right panel). 
           }
\end{figure}

Age calibrations using different combinations of the parameters 
$\zeta_\alpha$ and two different reference models are summarized in Table\,2.
The younger reference model is `Model S' 
\cite[(Christensen-Dalsgaard et al. 1996)]{jcd96} which has age 
$t_\odot=4.6\,$Gy; the second is `Model T', which has age
$t_\odot=4.7\,$Gy. The same physics was adopted in the
evolutionary calculations of both models. We notice in Table\,2 that 
the calibration for the combination $(A, C)$, i.e. without 
$\delta\gamma_1/\gamma_1$, is less 
stable to a change in the reference model than are the calibrations with 
combinations in which $\delta\gamma_1/\gamma_1$ is included, and therefore 
is less reliable, as we have explained in the introduction. If we ignore 
in Table\,2 the results for $(A, C)$ and combine the others, we obtain
\smallskip
\begin{center}
$t_\odot=4.68\,\pm0.02\,{\rm Gy}\,,\quad$
$Z=0.0169\,\pm0.0005$\,.
\end{center}
\smallskip
Including the calibrations with $(A, C)$ does not change the outcome.
Error contours corresponding to the calibration from Model S in the first 
row of Table\,2 are plotted in Figure\,3. Corresponding contours for Model T
are the same, except that their centres are displaced to (4.677\,Gy, 0.0170).
One can adduce from our description of the analysis in Section\,2 that our
current treatment of the errors is not completely unbiassed; however, the 
potential bias is of the order of only $|\delta\nu_i/\nu_i|$, which is small.

The age we have found is greater than currently accepted values. The values 
of $Z$ are somewhat smaller than those of Models S and T (0.01963), but we 
hasten to point out that they should not be 
regarded strictly as statements about the initial heavy-element abundance, 
but rather as measures of the opacity in the radiative interior. 
\cite{asplund04} have argued that the photospheric abundances of C, N 
and O had previously been overestimated, suggesting that the actual total 
heavy-element abundance is even lower than had previously been believed.  
However, 
that cannot imply that the opacity in the solar interior is necessarily 
comparably lower because it has been implicitly calibrated here (by accepting 
the tenets of solar-evolution theory, and the OPAL opacity calculations upon 
which the models are based), and indeed the opacity has already been 
determined seismologically from a broader spectrum of modes than has been 
adopted here (\cite[Gough 2004]{dog04}).  The matter raised by 
Asplund et\,al. therefore challenges either the opacity calculations, 
the nuclear reaction rates, or the basic physics of stellar evolution, not 
helioseismology as some spectators have surmised. As we 
know already from seismological structure inversions, the solar models are not 
accurate by helioseismological standards.  Therefore the properties inferred 
from these calibrations could be more contaminated by systematic error than by
errors in the observed frequencies.

\begin{table}
\centering
\caption{Age calibrations with different combinations of $\zeta_\alpha$ and
for the two reference models: Model\,S with an age $t_\odot=4.6\,$Gy 
and Model\,T with an age $t_\odot=4.7\,$Gy. The first two columns show the
results adopting Model S as the reference model, the third and fourth
columns display the results for Model T.
}
\begin{tabular}{lccccccc}
\noalign{\smallskip}
\noalign{\smallskip}
\hline
 \hfil$\zeta_\alpha$\hfil&
 \hfil $t_\odot$ (Gy)\hfil&
 \hfil$Z$\hfil&
 \hfil $t_\odot$ (Gy)\hfil&
 \hfil$Z$\hfil&
 \hfil$C^{1/2}_{\Theta 11}$\hfil&
 \hfil$-(-C_{\Theta 12})^{1/2}$\hfil&
 \hfil$C^{1/2}_{\Theta 22}$\hfil\\
\hline
$A,C,-\delta\gamma_1/\gamma_1$&4.679&0.0169&4.677&0.0170&0.017&-0.0023&0.0005\\
$A,C                         $&4.658&0.0177&4.673&0.0171&0.023&-0.0037&0.0007\\
$A,-\delta\gamma_1/\gamma_1  $&4.673&0.0165&4.676&0.0169&0.017&-0.0019&0.0007\\
$C,-\delta\gamma_1/\gamma_1  $&4.700&0.0169&4.680&0.0170&0.028&-0.0029&0.0005\\
\noalign{\smallskip}
\hline
\end{tabular}
\end{table}

\begin{figure}
\centering
\mbox{
\begin{minipage}[h]{63mm}
\includegraphics[height=.21\textheight]{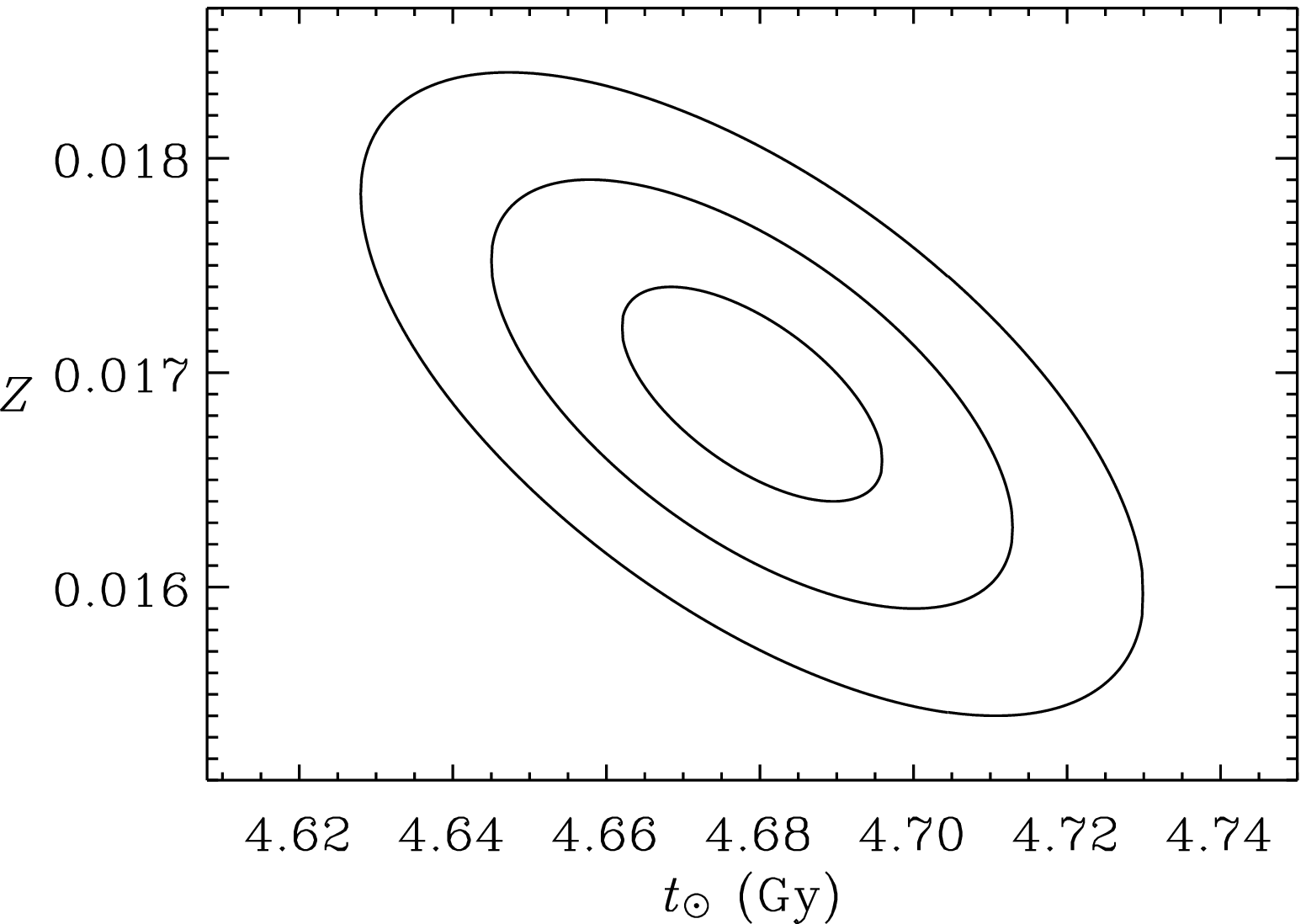}
\end{minipage}
\hspace{2mm}
\begin{minipage}[h]{64mm}
  \vspace{-24mm} 
  {{\normalfont\bfseries\upshape Figure 3.\,\ }
   Error ellipses for the calibration using all three parameters 
   $\zeta_\alpha$ and Models S as the reference model:
   solutions $(t_\odot,Z)$ satisfying the frequency data within 1, 2 and 3 
   standard errors in those data reside in the inner, intermediate 
   and outer ellipses, respectively.          
  }
\end{minipage}
}
\end{figure}

\begin{acknowledgments}
We thank J\o rgen Christensen-Dalsgaard for providing us with his 
stellar-evolutionary programme. GH acknowledges support by the 
STFC of the UK.
\end{acknowledgments}

\begin{discussion}
\discuss{Christensen-Dalsgaard}{A comment: with SONG we expect to be able
to carry out a similar analysis of distant stars, on which we of course know much less a priori.}
\discuss{Houdek}{This seismic diagnostic has been developed with the aim to be 
able to use it also for distant stars. The accuracy of the observed 
frequency data required for such a diagnostic analysis is one part in $10^4$ 
or better.}
\discuss{S. Vauclair}{Two small comments which are actually more relevant for
stars that are slightly more massive than the Sun; First, I would like to point 
out that in case of helium settling below the convective zone the effect of the
helium gradient in the second differences may become more important than the
convective border, and than the effect of helium ionization.
Second, the so-called asymptotic theory, which is very useful, may become 
quite wrong in some cases, at the end of the main sequence or the beginning of
the sub-giant branch. The small frequency separation, which is always positive
in the asymptotic theory, can become negative.
} 

\discuss{Gough \& Houdek}{You are certainly correct in implying that the
amplitude of the oscillatory contribution to the second differences arising
from helium settling beneath the convection zone can be greater in stars more
massive than the Sun, which have shallower convection zones, although whether
or not it is more important than the ionization signature depends upon the
issue in question. The cumulative amount of settling increases with time, and
therefore is a potential indicator of age. But the sound-speed profile that it
produces depends on uncertain fluid-dynamical issues associated with the
tachocline, the recession of the convection zone, and possible overshooting, so
we would be wary of attempting to use its seismic signature in an age 
calibration. In the current state of our understanding we would instead prefer
to separate it from the ionization signature and then ignore it, as we have 
for the Sun; that course is possible provided that the helium ionization zone
is acoustically far from the base of the convection zone. We would use it
separately to investigate tachocline structure, however, as indeed we are 
in the process of doing for the Sun.

One cannot deny that conditions in some stars might be such as to render it 
impossible to develop an adequate asymptotic theory of low-degree acoustic 
modes, although we do not share your apparently implied pessimism. It is
perhaps worth pointing out that there have been instances when an asymptotic
formula developed in one set of circumstances has been misused by applying
it without modification in another, in which the conditions for the validity
of the theory are not satisfied; if that is what you mean by ``so-called
asymptotic theory'' we must surely agree. We must point out, however, that it
is not true that even the simple asymptotic glitch-free formula 
(\ref{eq:asymp}) precludes a negative so-called small frequency separation.
}
\end{discussion}

\vspace{-3mm}
\begin{thebibliography}{}
\bibitem[Asplund et al. (2004)]{asplund04}
{Asplund M., Grevesse N., Sauval A.~J., Allende Prieto C., 
Kiselman D.} 2004,  \textit{A\&A} 417, 751
\bibitem[Basu et al. (2007)]{basu07}
  {Basu S., Chaplin W.~J., Elsworth Y., New A.~M., Serenelli G.,
   Verner G.~A.} 2007, \textit{ApJ} 655, 660
\bibitem[Bonanno, Schlattl \& Patern\`o (2002)]
        {bon-schlat-pat02}
{Bonanno A., Schlattl H., Patern\`o L.} 2002, \textit{A\&A} 390, 1115
\bibitem[Christensen-Dalsgaard (1982)]{jcd82}
{Christensen-Dalsgaard J.} 1982, \textit{MNRAS} 199, 735
\bibitem[Christensen-Dalsgaard (1984)]{jcd84}
{Christensen-Dalsgaard J.} 1984, in: Mangeney A., Praderie F., (eds), 
\textit{Space Research Prospects in Stellar Activity and Variability}, 
Paris Observatory Press, Paris, p.\,11
\bibitem[Christensen-Dalsgaard (1988)]{jcd88}
{Christensen-Dalsgaard J.} 1988, in: Christensen-Dalsgaard J., Frandsen S., 
(eds), \textit{Proc. IAU Symp.\,123, Advances in helio- and asteroseismology},
Reidel, Dordrecht, p.\,295
\bibitem[Christensen-Dalsgaard et al. (1996)]{jcd96}
{Christensen-Dalsgaard J. et al.} 1996, \textit{Sci} 272, 1286 
\bibitem[Dziembowski et al. (1999)]{wd99}
{Dziembowski W.~A., Fiorentini G., Ricci B., Sienkiewicz R.} 1999, 
\textit{A\&A} 343, 990
\bibitem[Gough (1983)]{dog83}
{Gough D.~O.} 1983, in: Shaver P.A., Kunth D., Kj{\"a}r K., (eds),
\textit{Primordial helium}, Southern Observatory, p.\,117
\bibitem[Gough (1987)]{dog87}
{Gough D.~O.} 1987, \textit{Nat.} 326, 257
\bibitem[Gough (2001)]{dog01}
{Gough D.~O.} 2001, in: von Hippel T., Simpson C., Manset N., (eds),
\textit{ASP Conf. Ser. Vol.\,245, Astrophysical ages and timescales}, 
{Gough D.~O.} 1987, \textit{Nat.} 326, 257
\bibitem[Gough (2002)]{dog02}
{Gough D.~O.} 2002, in: Favata F., Roxburgh I.W., Gadal\'i-Enr\'iquez D., 
(eds), \textit{Proc 1st Eddington Workshop: Stellar structure and 
habitable planet finding}, ESA SP-485, Noordwijk, p.\,65
\bibitem[Gough (2004)]{dog04}
{Gough D.~O.} 2004, in: {\v C}elebonovi\'c V., D\"appen W., Gough D.~O., (eds), 
\textit{AIP Conf. Proc. Vol.\,731, Equation-of-state and phase-transition 
issues in models of ordinary astrophysical matter}, 
Am. Inst. Phys., Melville,\,p.119
\bibitem[Gough \& Novotny (1990)]{dog-nov90}
{Gough D.~O., Novotny E.} 1990, \textit{Solar Phys.}, 128, 143
\bibitem[Guenther (1989)]{guenther89}
{Guenther D.~B.} 1989, \textit{ApJ} 339, 1156
\bibitem[Guenther \& Demarque (1997)]
{guenther-demarque97}
{Guenther D.~B., Demarque P.} 1997, \textit{ApJ} 484, 937
\bibitem[Houdek \& Gough (2007a)]{hgdog07a}
{Houdek G., Gough D.~O.} 2007a, \textit{MNRAS} 375, 861
\bibitem[Houdek \& Gough (2007b)]{hgdog07b}
{Houdek G., Gough D.~O.} 2007b, in: Stancliffe R.J., Dewi J., Houdek G., 
Martin R.G., Tout C.A., (eds), \textit{AIP Conf. Proc.: Unsolved Problems 
in Stellar Physics}, American Institute of Physics, New York, p.~219
\bibitem[Ulrich (1986)]{ulrich86}
{Ulrich R.~K.} 1986, \textit{ApJ} 306, L37
\bibitem[Weiss \& Schlattl (1998)]{weiss-schlattl98}
{Weiss A., Schlattl H.} 1998, \textit{A\&A} 332, 215

\end{thebibliography}
\end{document}